PHYSICS

# How crystals form: A theory of nucleation pathways

James F. Lutsko

Recent advances in classical density functional theory are combined with stochastic process theory and rare event techniques to formulate a theoretical description of nucleation, including crystallization, that can predict nonclassical nucleation pathways based on no input other than the interaction potential of the particles making up the system. The theory is formulated directly in terms of the density field, thus forgoing the need to define collective variables. It is illustrated by application to diffusion-limited nucleation of macromolecules in solution for both liquid-liquid separation and crystallization. Both involve nonclassical pathways with crystallization, in particular, proceeding by a two-step mechanism consisting of the formation of a dense-solution droplet followed by ordering originating at the core of the droplet. Furthermore, during the ordering, the free-energy surface shows shallow minima associated with the freezing of liquid into solid shells, which may shed light on the widely observed metastability of nanoscale clusters.

## INTRODUCTION

Crystallization has been called "one of the most secretive processes in nature" (*1*). The mystery lies in how structures with long-range order form from building blocks that only interact with their local neighbors. The most extreme case is when solids form from weak solutions, in which the molecules begin widely separated so that not even local structure is present. Nucleation, in general, and crystallization, in particular, attract a great deal of interest because of their widespread practical importance in many biological and technical contexts. Liquid-liquid phase separation is now recognized as "underlying the formation of several membraneless compartments in living cells, including, for example, stress granules, the nucleolus, and P bodies in the cytoplasm" (*2*). Crystallization plays a role in several pathologies such as malaria (*3*) and amyloid aggregation leading to Alzheimer's disease (*4, 5*), while the control of polymorphism in industrial processes, such as the production of pharmaceuticals (*6*), and at the nanoscale, where surface effects enhance polymorphism (*7, 8*), is a subject of intense practical interest.

One of the most important insights from the recent work has been the discovery of nonclassical pathways for crystallization. The naive picture of crystallization is that a few molecules come together by chance and happen to be arranged in a crystalline form. Additional molecules attach one by one, gradually building a larger structure. Small clusters are unstable because most molecules are near the surface and do not have the correct number of neighbors, but large enough clusters are stable. The same picture holds for liquid-liquid separation except, in that case, the clusters are disorganized droplets. Remarkable progress in both the observation of crystallization processes—via techniques such as cryogenic transmission electron microscopy (cryo-TEM) (*9, 10*) and atomic electron tomography (*11*)—and simulation using rare event techniques (*12*) has been made in recent years, and it is now recognized that the process of reaching a crystalline critical cluster is often much more complicated than supposed in the naive scenario, with multiple intermediate phases sometimes playing a role in reaching the final state [see, e.g., (*12*) for a recent review of simulation studies and (*13*) for recent developments with water]. The discovery of the "two-step" mechanism of crystallization in proteins has been an important factor in driving interest in the subject (*14, 15*). Since then, complex pathways have been reported for many other systems including calcium carbonate (*16, 17*), potassium dihydrogen phosphate (*18*), polyoxometalates (*10*), and biomimetic polymers (*19*), to name a few. As a result, nucleation pathways have become a central focus of modern research.

It is therefore surprising that there is no fundamental theory of nucleation that can explain, much less predict, nucleation pathways (and energy barriers and rates) from first principles. While there have long been rules of thumb concerning metastable intermediate states, such as Ostwald's rule of stages and the Stranski-Totomanow conjecture, they do not take into account the kinetics that can dominate the selection of the pathways (*20*). In the following, the elements of such a theory, based only on the interaction potential of the molecules, are presented and applied to both liquid-liquid separation and crystallization of colloids and macromolecules in solution.

## Theory

Any theory of nucleation requires two elements: first, a means of calculating properties of clusters such as their structure and free energies, and second, a dynamical description of fluctuations. Classical density functional theory (cDFT) (*21, 22*) has long been recognized to have the potential to provide the first element. Since the pioneering work of Oxtoby and Evans in the 1980s (*23*), cDFT has been used to determine the structure and energy of critical clusters for, first, nucleation of liquid droplets from a vapor and, later, that of crystallization. Early calculations involved many simplifying assumptions, but more recent work has demonstrated quantitative agreement with simulation (*24*). The advantage of cDFT over alternative techniques such as phase-field crystal (PFC) theory (*25*) and diffuse interface models is that it is, in principle, "ab initio," requiring only an interatomic potential as input and that it gives a quantitatively accurate description of molecular-scale correlations and structure. cDFT is a fundamental theory from which others can be understood as approximations (*22*).

In the implementation of cDFT used here, the local number density of each chemical species is discretized on a cubic computational lattice, with lattice spacing much smaller than the individual molecules. The equilibrium local densities are determined by minimizing a functional of the local densities, giving both the equilibrium distribution of the molecules and the free energy of the system. The theory correctly describes molecular-scale features such as packing of dense fluids into layers near a wall. In particular, while a homogeneous fluid has a uniform density, a solid is intrinsically nonuniform at the molecular

Center for Nonlinear Phenomena and Complex Systems, Université Libre de Bruxelles, Code Postal 231, Blvd. du Triomphe, 1050 Brussels, Belgium.
Email: jlutsko@ulb.ac.be

scale as the density is sharply peaked at the lattice sites and goes to very low values between them. Recent advances in cDFT have extended its applicability to highly nonuniform systems such as dense liquid droplets and solid clusters in equilibrium with a background of low-density vapor (see Fig. 1).

The second element necessary to describe nucleation is a description of fluctuations (26). A natural framework for this is fluctuating hydrodynamics (FH), which was originated by Landau and has been intensively studied and developed. FH is now a widely used tool that has been applied to a range of subjects such as mode coupling theory, glass transition, and nucleation, and its foundations in more fundamental statistical mechanics have been established [see, e.g., (27)]. The basic quantities used in the theory are the spatially varying local densities of each species as well as the velocity and temperature fields. For large particles, such as colloids or macromolecules, in a bath of smaller particles (e.g., water), an approximate effective description of the large molecules can be derived in which the effect of the smaller molecules is modeled as a combination of friction and a stochastic force. If the damping of the bath is strong, then this can be further reduced to a single equation describing the density field for the large species (28, 29) having the form

$$\frac{\partial}{\partial t}\hat{n}_t(\mathbf{r}) = D\nabla \cdot \hat{n}_t(\mathbf{r})\nabla \frac{\delta F[\hat{n}_t]}{\delta \hat{n}_t}(\mathbf{r}) + \nabla \cdot \sqrt{2D\hat{n}_t(\mathbf{r})}\hat{\xi}_t(\mathbf{r}) \quad (1)$$

where $\hat{n}_t(\mathbf{r})$ is the fluctuating local density that is a nonequilibrium quantity: The local density of cDFT is the fluctuation-averaged value of this for an equilibrium system. The coefficient $D$ is the coefficient of diffusion in the low-density limit (i.e., when there is only a single large molecule in the bath undergoing Brownian motion), which can be calculated given the properties of the molecules making up the system. The free-energy functional $F[\hat{n}_t]$ is taken to be the Helmholtz functional of cDFT. At the level of hydrodynamics (before assuming the overdamped limit), this is more generally related to the gradient of the pressure in the FH equations, and its use is a kind of local equilibrium approximation common in nonequilibrium statistical mechanics (26). Last, $\hat{\xi}_t(\mathbf{r})$ is the local white noise (with delta-function correlations in space and time) arising from the small bath molecules colliding with the large molecules and is the origin of fluctuations in the model. An important property of this model is that it conserves the number of particles at all times, except possibly at the borders of the system. Here, I concentrate on this simple, yet realistic, model and leave similar developments under less restrictive assumptions to future work.

The use of the cDFT free-energy functionals in stochastic models has been questioned (30) because what occurs in typical derivations of the stochastic models is a coarse-grained free energy and not the equilibrium cDFT functional. The difference is due to the fluctuations explicitly represented in the stochastic model by the fluctuating force, and it is expected that the cDFT functional results from a fluctuation-average of the coarse-grained functional. Here, as in almost all applications, the free-energy functional is taken to be the sum of a sophisticated hard-sphere functional and a mean-field treatment of the attractive tail of the potential (22). For the hard-sphere contributions, such an average can be expected to have little effect because all correlations are short ranged. For systems with long-range attractive tails, the mean-field description used here and in all similar applications can be expected to be more justifiable for the coarse-grained models than for the fluctuation-averaged model because this averaging is precisely the physical origin of renormalization effects that invalidate the mean-field description, e.g., in critical phenomena. Hence, the combination of a cDFT hard-sphere free-energy functional and a mean-field attractive tail could be argued to be a good guess at a coarse-grained free energy and a bad guess at a cDFT functional, rather than the reverse.

It would be possible to select one of the cDFT model functionals for use in the stochastic model and to perform direct numerical simulations. This approach has recently been exploited and is a promising coarse-grained simulation method to study nucleation (31). Another possibility is to further coarse-grain via the introduction of collective variables or order parameters. This route has been explored in other places, where it has been shown that one can recover classical nucleation theory (CNT) with appropriate approximations (26, 32). Thus, the theory discussed here is not an alternative to CNT but rather a more fundamental theory for which CNT is an approximation. Here, however, the aim is to continue the theoretical development using tools from the theory of stochastic processes and to concentrate on the nucleation pathway as the fundamental object.

If a system begins as a weak solution (i.e., a vapor-like state) and spontaneously nucleates a cluster (either a droplet of dense solution or a crystalline cluster), then the initial local concentration is constant throughout the system. When the cluster is present, the concentration is high inside the droplet and low outside: The difference between the states can be characterized entirely in terms of the local density. For stochastic models such as the one used here, it is possible to give an exact expression for the probability to follow any given path from some initial density distribution, $n_0(\mathbf{r})$, to any given final one, $n_T(\mathbf{r})$, and by searching for the path with maximum probability, one can determine the most likely path (MLP) that characterizes the transition. The basic idea goes back to Onsager and Machlup (33), and the generalization to arbitrary diffusive processes was given by Graham (34). If the initial state is the uniform mother phase and the final state includes the critical (or post-critical) cluster of the new phase, then the MLP will be the most likely nucleation pathway. In general, determining the MLP is highly nontrivial, but important simplifications take place in the weak noise limit (corresponding physically to, e.g., lower temperatures), in which case the stochastic theory is equivalent to the Wentzell-Freidlin large deviation theory (35). Then, one can prove (26) that the MLP for nucleation must pass through the critical cluster—a fact that is not generally true in the strong noise limit and shows that the usual picture of nucleation only applies in this limit. Furthermore, one can show that the MLP can be constructed by starting at the critical cluster and perturbing the system slightly in the unstable directions so that the deterministic part of the dynamics

$$\frac{\partial}{\partial t}\hat{n}_t(\mathbf{r}) = D\nabla\cdot\hat{n}_t(\mathbf{r})\nabla\frac{\delta F[\hat{n}_t]}{\delta\hat{n}_t(\mathbf{r})} \qquad (2)$$

causes the density to fall down the free-energy gradient to either the initial phase or the final phase, depending on the direction of the perturbation. Putting these two partial paths together gives the complete MLP. Note, however, that they have two very different characters: In reality, the system starts in the initial phase and is then driven by fluctuations up the free-energy barrier until it reaches the critical cluster, after which it then continues to grow until as much material as possible is incorporated into the new phase. The second part, starting at the critical cluster and growing, is only a normal thermodynamically driven growth driven by the free-energy gradient. The first part, however, is driven against the free-energy gradient by fluctuations and it is a highly nontrivial and useful result that the MLP for this process can be determined by falling "backward" down the gradient (26).

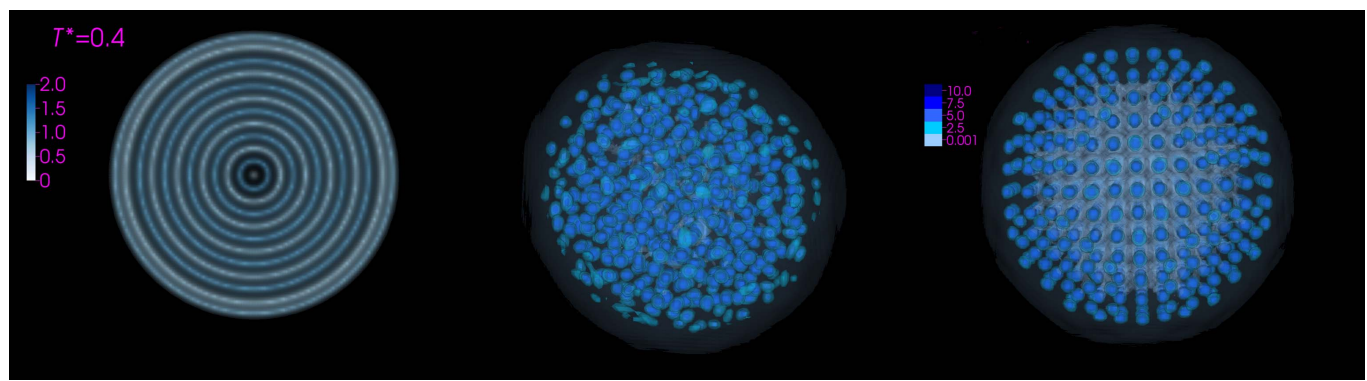

**Fig. 1. Typical structures obtained from three-dimensional cDFT calculations, as reported in (50).** Each figure is a view of the local density obtained by free-energy minimization of the Lennard-Jones system at a reduced temperature of $T^* = 0.4$. The leftmost figure is a slice through a dense-solution, liquid-like, droplet; the center figure is a contour representation of an amorphous, glass-like cluster; and the rightmost figure is a contour representation of a face-centered cubic (fcc) cluster. The droplet shows the packing into shells separated by low-density regions, which is typical of confined liquids. In the other two structures, the density is localized into "atoms" separated by very low density regions.

Rather than directly using gradient descent, i.e., Eq. 2, from critical clusters, the present work makes use of the string method (36), which is mathematically equivalent but offers advantages of efficiency and simplicity. In particular, it determines the entire pathway at once and is particularly useful when there are multiple intermediate free-energy minima or when the free-energy gradients are weak—both of which prove relevant below. Details of the implementation are discussed in Supplementary Text, and here, it is only noted that, in the string method, one works with a collection of densities, or "images," distributed all along the pathway, thus approximating it as a collection of discrete points. By moving them according to Eq. 2 under the constraint of maintaining the equal spacing between points, the gradient-descent pathway is determined. The method requires an initial guess for the pathway, and for this, a simple linear interpolation between the end points was used. In the present calculations, the starting point, i.e. the initial image, is the uniform, low-density system, and the final point on the pathway is the critical cluster. The calculations reveal how the system evolves from the former to the latter.

It is appropriate to mention that the individual elements of this theory have been discussed previously in similar contexts. For example, in the study of Lutsko (24), the very similar nudged elastic band method was used together with the state-of-the-art cDFT free-energy functional to describe liquid-liquid nucleation. However, in that work, the importance of introducing a realistic dynamical description was not understood. Similarly, phase-field studies such as those of Qiu and Qian (37) and Backofen and Voigt (38, 39) use simpler free-energy functionals and ad hoc dynamics together with the string method. While the critical clusters are correctly determined, the physicality of the pathways is unclear because of the abstract nature of the dynamics (e.g., the lack of local mass conservation when the order parameter is to be interpreted as the density). In addition, the excluded volume effects, which dominate the structure at the molecular level, are outside the domain of these models. The present work uses the sophisticated fundamental measure theory model for the hard-sphere contribution to the free-energy functional, which is well known to give a very accurate description of hard-sphere systems in particular [see, e.g., (40)] and, when combined with a mean-field model for the attractive part of the potential, of molecular-scale structure for more general potentials (22).

Approaches to crystallization very close in spirit to the one presented here have also been explored for some time in the PFC community (25). These can be understood as cDFT models that are simplified by expanding the free-energy functional $F[n]$ about a uniform state, $n_0$, resulting in two types of terms (41, 42). The first set of terms takes the form of a gradient expansion, which is truncated at fourth order. The second set of terms takes the form of an expansion in the variable $\phi(\mathbf{r}) \equiv (n(\mathbf{r}) - n_0)/n_0$, although this expansion is sometimes avoided (43). Both expansions are uncontrolled in the solid state, where variations of the density of several orders of magnitude occur over the smallest molecular length scales, and for this reason, PFC models cannot generally be parameterized to a particular interaction potential. Rather, the effective potential underlying them is determined by the approximations used and can be highly unconventional [see, e.g., (44)]. This simplified free-energy functional has been coupled to dynamics in many studies, including some in which the combination is very similar to that used here (45), as a basis for coarse-graining of the propagation of crystallization fronts in two dimensions (43) and of nucleation (44–46). The aim of the present work is to overcome the limitations of these simplified models while focusing on the nucleation pathway via the concept of the MLP.

## RESULTS

Fully three-dimensional calculations are reported here for particles interacting via a Lennard-Jones potential (see Supplementary Text for details of the calculations). The system has a typical phase diagram with a weak liquid, a dense liquid, and a crystalline phase (see fig. S1). Calculations were performed to determine the nucleation pathways for both liquid-liquid phase separation and crystallization for the case of open systems so that the process takes place at constant chemical potential and with a variable number of molecules in the calculational cell. Note that, because mass is locally conserved by the dynamics, matter can only enter and exit the cell via its boundaries. The energetics of the process is therefore presented below in terms of the grand canonical free energy, $\Omega = F - \mu N$, where $F$ is the canonical or Helmholtz free energy, $\mu$ is the chemical potential, and $N$ is the total number of molecules in the calculational cell. For both droplet nucleation and crystallization, a variety of systems were studied, with critical nuclei ranging from 100 to 400 molecules. In the following, results are described for typical calculations selected from many that have been performed. The Supplementary Text gives details of many more results as well as tests of the stability of the method with respect to the number of images used and the demonstration of the robustness of the results obtained from very different initial conditions.

### Liquid-liquid separation

The simplest phase transition is a liquid-liquid, or weak-solution to dense-solution, transition that is analogous to a vapor-liquid transition in a simple fluid and that, aside from its simplicity, is an important process in its own right, as discussed above. The free energy along the nucleation pathway for conditions that correspond to a supersaturation of $S = 3.2$ is shown in Fig. 2. The prediction of CNT calculated using input (e.g., the coexistence of free energies of the two phases and the surface tension) obtained from separate cDFT calculations, as well as a variation in which a nonzero Tolman length is included, is also displayed. The CNT predictions, with no adjustable parameters, are qualitatively reasonable, and fitting to the Tolman length gives good quantitative agreement once the cluster becomes sufficiently large, which is not surprising given the assumptions of CNT. However, while in CNT the nucleating cluster is viewed as beginning with zero radius and growing monotonically, the inset shows that the process begins with a small increase in density over a large volume—a long-wavelength, small-amplitude density fluctuation—which translates into a cluster with a small density but a large radius. This becomes smaller and denser until the cluster becomes liquid-like, at which point it begins to grow along something close to the CNT pathway. These results extend those obtained previously with more coarse-grained models (47, 48) that imposed spherical symmetry to fully unconstrained, three-dimensional models that include molecular-scale correlations. They are also broadly consistent with earlier work, suggesting the failure of CNT to describe the early stages of nucleation (49). Figures 3 and 4 show snapshots of the developing cluster; movies showing the complete sequence are included in the Supplementary Materials. While the marked difference from CNT may appear surprising, it is quite natural considering that mass is locally conserved during nucleation. Thus, to form a dense droplet, an excess of matter must come together via thermal fluctuations. The only question then is whether mass somehow is supplied "on demand" as a cluster grows or whether a cluster forms in regions where the local mass density is already high. The present results show that the latter is more probable as one might guess: Nucleation is more likely in regions of (locally) higher supersaturation.

### Crystallization

To study crystallization in an open system, a metastable crystalline cluster was constructed using a method from the literature (50) (details are given in the Supplementary Text). This was then used as an end point for the computation of a path beginning with the uniform, weak solution. The results for several different supersaturations are largely similar. A particular example is shown in Figs. 5 and 6, and corresponding movies are provided in the Supplementary Materials. In broad outline, the two-step mechanism for nucleation is immediately apparent. Just as in the case of liquid-liquid separation, the process starts with a small increase of the background concentration over a large region, i.e., a long-wavelength, small-amplitude density fluctuation. From this, a high-concentration droplet forms and densifies, showing increasing structure as it grows in size. Around the 20th image of the sequence of 50 used in the calculation, the dense interior begins to show localization into a few "molecular" sites. The localization spreads with virtually no change in the number of molecules until the crystalline state is reached, at which point growth resumes. Crystallization begins at the center of the cluster and spreads outward. The dense droplet created in the first part of the process is always subcritical, and therefore unstable, as the critical droplet for this supersaturation contains about 350 molecules (see Supplementary Text). Thus, while the nucleation pathway begins by the formation of droplets, the solid does not form via a nucleation event within a metastable, growing droplet as is sometimes imagined. These results are typical of calculations performed for supersaturations from 2.8 to 4.3 (150 to 400 molecules in the critical cluster).

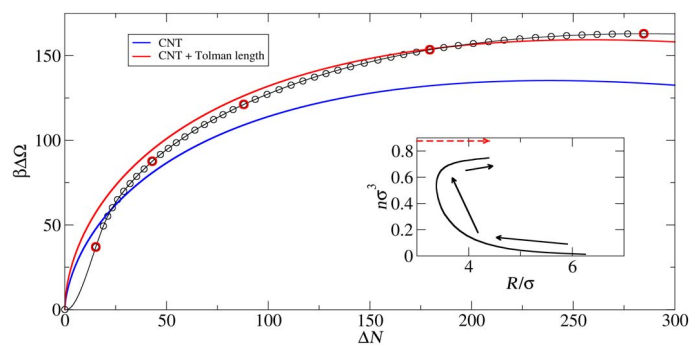

**Fig. 2. Droplet free-energies as a function of size as determined from CNT and the present theory.** Nucleation pathway for a cluster of 286 molecules at $k_B T = 0.475\varepsilon$ and supersaturation $S = 2.2$, displayed as the excess grand canonical free energy (i.e., relative to the weak-solution free energy) versus the excess number (relative to the weak solution) of particles in the simulation cell (which is effectively the number of particles in the cluster). The blue line is the prediction of CNT, and the red line is the result of allowing the surface tension ($\gamma$) to depend on the radius ($R$) of the cluster [$\gamma = \gamma_0(1 + l/R)$] and fitting to the larger clusters (those with $\Delta N > 100$) with the result that the Tolman length $l = 0.232\sigma$. The images marked in red are shown in subsequent figures. The inset shows the cluster radius (based on the radius of gyration, as described in the text) and the average density for particles within the sphere of radius $R$. The arrows indicate the direction of movement along the MLP. The broken line is the CNT path.

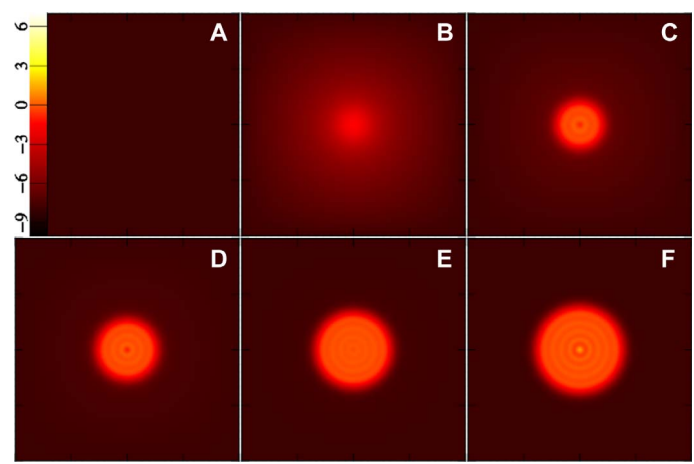

**Fig. 3. Snapshots of slices through the strong-solution clusters along the nucleation pathway, leading to the critical cluster composed of about 286 molecules.** The images show the density on a logarithmic color scale ranging from black [$\ln(n\sigma^3) = -10$ or lower] to red to yellow to white [$\ln(n\sigma^3) = 7$ or higher]. (**A**) Point at the origin of Fig. 2, i.e., the pure weak solution. Counting from that one, (**B**) to (**F**) correspond to points 1, 10, 20, 35, and 49, respectively, with the latter being the final point, i.e., the critical cluster.

Upon closer examination (see the inset in Fig. 5), one sees that the free energy during the crystal growth stage is not monotonic but rather shows a shallow minimum. The figures show that the energy is nearly constant during the initial ordering but that it then increases as a new layer of crystal forms on the outside of the cluster. This is not surprising: It is well known that the energy of crystalline clusters manifests minima at "magic" numbers of growth units corresponding to completed crystalline shells, and their effect on the nucleation rate has been the subject of recent research (51, 52). The nonmonoticity observed here is a similar phenomenon but differs in detail. In the CNT picture assumed in the previous work, molecules are added one by one so that the cluster must pass through unfavorable "nonmagic number" intermediate states: Here, a dense-solution (liquid-like) layer forms on the surface of the crystal and freezes into multiple localized sites at once so that the energy barrier is associated with the freezing transition rather than the addition of individual growth units. In general, this can be expected to lead to slower nucleation rates due to the multiple barriers that must be overcome. Furthermore, if there are many shallow valleys that must be traversed to reach the post-critical state, then the clusters can be expected to spend a considerable time in this "desert" with no strong driving force to either grow or dissolve. It is tempting to speculate that this could provide a mechanism for the metastability of the mysterious precritical clusters that have been observed in many different systems [see, e.g., (3, 53, 54)].

## DISCUSSION

It has been shown in recent years that this theory, which has been called mesoscopic nucleation theory, is not an alternative to the familiar CNT but rather is a more fundamental description from which CNT can be systematically derived (26, 32, 48). In principle, it requires no input other than the interaction potential. Of course, the quality of the results depends on the cDFT and dynamical model for fluctuations. Here, I have used a simple but widely accepted model for the dynamics, but there is much work on extensions to include, e.g., hydrodynamic-mediated interactions between the colloidal particles (55). An important extension will be the description of nucleation in single-component systems, which will require the use of the full FH framework, including heat transport, rather than the overdamped limit used here.

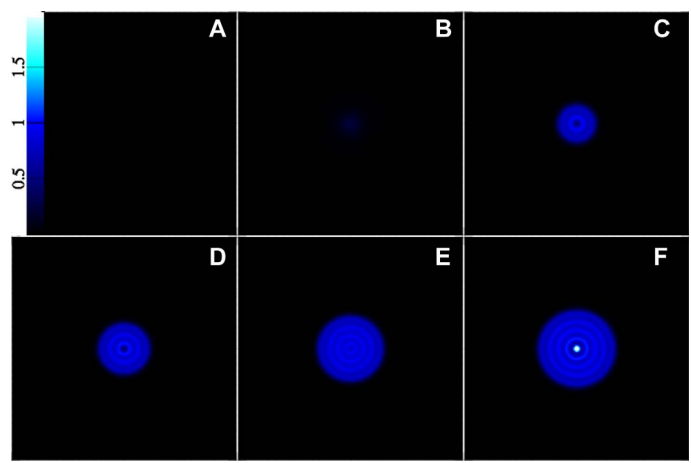

**Fig. 4. Snapshots of slices through the strong-solution clusters, as in Fig. 3 but showing the density on a linear color scale ranging from black ($n\sigma^3 = 0$) to dark blue and light blue to white ($n\sigma^3 \geq 2$).** The packing structure, which manifests as alternating spherical shells of high and low density, is clearly visible in this view.

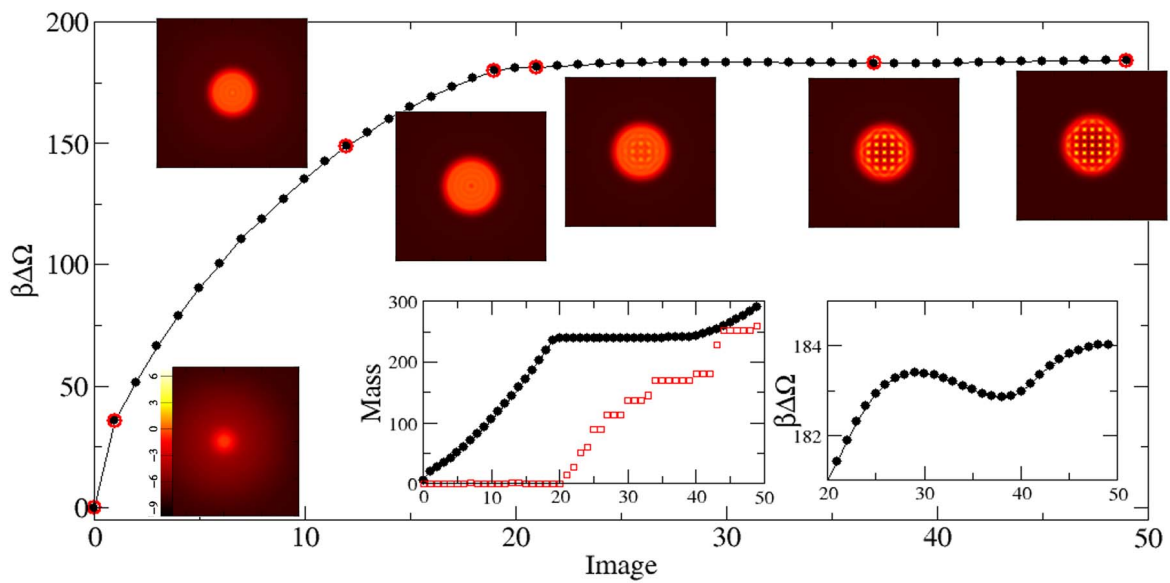

**Fig. 5. Nucleation pathway for a cluster of approximately 286 growth units:** The lines and points plot the excess free energy along the path as a function of an abstract reaction coordinate. Inset images show the log of the density on a slice through the center of the cluster for the free energy points outlined in red. The inset graph on the left shows the mass of the cluster (black symbols) and the number of density peaks above a threshold of $n\sigma^3 > 5$ (red symbols), which is a measure of the number of "solid" molecules; the one on the right is a magnified look at the free energy in the latter stages of the process. The path displays the two-step mechanism consisting of the formation of a droplet followed by freezing at constant mass. As shown in the magnified view, the free energy also contains a shallow minimum after the initial barrier. An examination of the images shows that the minimum is due to the completion of a crystalline shell. The free energy barrier is about 175 $k_BT$.

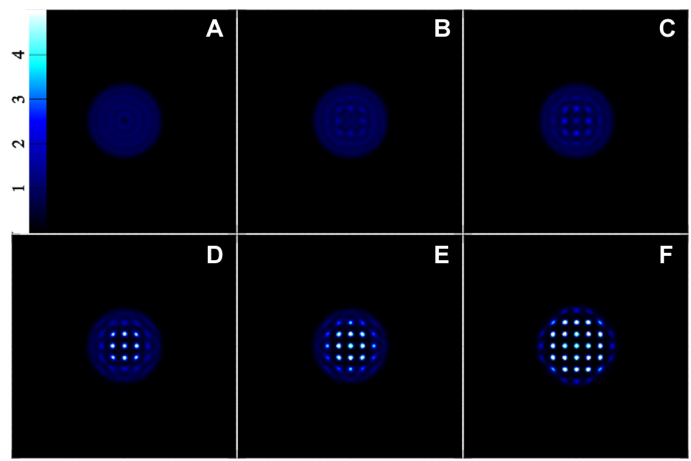

**Fig. 6. Snapshots of the density on a linear scale during the onset of crystallization.** Panels (**A**) to (**C**) correspond to points 19, 20, and 21, respectively, on Fig. 5. The later stages of crystallization are shown in panels (**D**) to (**F**) corresponding to points 25, 30, and 49, respectively, on Fig. 5.

The work described has several connections to previous work on crystallization. The structures shown above for the early onset of solid-like ordering are similar to the "onion-structure" model for the critical nucleus proposed by Barros and Klein (*56*), and it could be useful to analyze the structures here to make quantitative comparison to their work. Several PFC simulations have demonstrated crystallization precursors—amorphous structures were reported by Tóth *et al.* (*44*) and Tang *et al.* (*46*) saw more complicated scenarios involving both medium-range liquid-order and body-centered cubic (bcc) precursors to face-centered cubic (fcc) crystals. In a separate study, Tóth *et al.* (*45*) also noted multiple similar free-energy minima in their PFC free-energy landscapes, which is suggestive of the metastable structures reported here.

The methods demonstrated here for homogeneous nucleation translate effortlessly to other problems such as nucleation in closed systems, heterogeneous nucleation, mixtures with chemical reactions, wetting, and many others. For example, heterogeneous nucleation is only more complicated in that specific interactions and geometries of surfaces or seeds must be specified, and the equivalent of the critical clusters shown in Fig. 1 for the case of solutions in contact with a hydrophobic wall has already been presented (*50*). In principle, nucleation rates can be calculated using the classical theory developed by Langer (*57*), Langer and Turski (*58*), Hänggi *et al.* (*59*), Talkner (*60*), and others, given the free-energy surfaces and gradient dynamics. In the case of crystallization, if there are multiple shallow metastable states leading to long-lived intermediaries, then the classical theories will only be directly applicable to the first large nucleation barrier and the determination of the entire rate will depend on new theoretical developments that will require further characterization of the complex free-energy surfaces that has only begun here. For example, if the free-energy surface is almost flat, then recrossing the barrier, and falling back to the initial state, will be much more common than for parabolic barriers and will lead to smaller nucleation rates. A simple exploration of this effect for a one-dimensional system is given in the Supplementary Text, where the conclusion is that a plateau changes the kinetic prefactor, but not the dominant exponential factor, of the nucleation rate. In any case, the prediction of multiple metastable intermediate states arising during crystallization—both smaller and larger than the critical cluster—suggests a new possibility for understanding metastable clusters often observed in experiment (*3*, *53*, *54*).

## MATERIALS AND METHODS
### The MLP
For generalized diffusion processes such as those used here (Eq. 1), the probability to observe any given path $n_t(\mathbf{r})$ for $0 \leq t \leq T$ with fixed end points $n_0(\mathbf{r})$ and $n_T(\mathbf{r})$ was given by Graham (*34*) as

$$P[n_t] \sim \exp\left(-\int_0^T L[\dot{n}_t, n_t] dt\right) \quad (3)$$

with $\dot{n}_t = d\dot{n}_t/dt$, where the explicit form of the Lagrangian, $L[\dot{n}_t, n_t]$, is rather complex in the general case [see, e.g., (*26*, *34*) for the explicit expression]. The MLP is the function $n_t(\mathbf{r})$ that connects the specified end points and maximizes this probability. In the weak noise limit, the Lagrangian may be adequately approximated by

$$L[\dot{n}_t, n_t] \simeq \int \left(\dot{n}_t(\mathbf{r}) - \nabla \cdot n_t(\mathbf{r}) \nabla \frac{\delta F[n_t]}{\delta n_t(\mathbf{r})}\right)(\nabla \cdot n_t(\mathbf{r})\nabla)^{-1}$$
$$\left(\dot{n}_t(\mathbf{r}) - \nabla \cdot n_t(\mathbf{r}) \nabla \frac{\delta F[n_t]}{\delta n_t(\mathbf{r})}\right) d\mathbf{r} \quad (4)$$

and it can be shown (*26*, *61*) that if the end points are separated by an energy barrier, then the MLP determined in this approximation passes through the critical point separating them and is determined by

$$\frac{\partial}{\partial t} \hat{n}_t(\mathbf{r}) = \pm D \nabla \cdot \hat{n}_t(\mathbf{r}) \nabla \frac{\delta F[\hat{n}_t]}{\delta \hat{n}_t(\mathbf{r})} \quad (5)$$

with the sign chosen according to the direction one wishes to move along the path. In this case, the variable $t$ is not a physical time but serves to parameterize the geometric path. Note that this is a type of gradient descent with the metric $(\nabla \cdot n_t \nabla)^{-1}$ in density space.

### The string method: Relaxation
The string method alternates relaxation steps and reparameterization steps operating on a set of $N + 1$ images $n_j(\mathbf{r})$ along the path with $n_0 = n_{t=0}$ and $n_N = n_{t=T}$. The relaxation step consists of integrating each image for a short time $\delta t$ according to the dynamics given in Eq. 5 except for the end points, $j = 0,N$, which are held fixed. When the density is sharply peaked, as in the solid phase, these equations are very stiff. The algorithm used here consists of the exact integration of the purely diffusive part of the dynamics arising exactly from the ideal part of the free energy and a semi-implicit scheme for the part coming from the non-ideal free energy—all taking into account the discrete nature of the problem. The detailed algorithm is given in the Supplementary Text.

### The string method: Reparameterization
The second part of the string method requires using the relaxed images to determine new, evenly spaced images along the path. The exact manner in which "evenly spaced" is defined is unimportant, and here, the Euclidean distance, $d[n_j, n_{j+1}] = \sqrt{\int (n_{j+1}(\mathbf{r}) - n_j(\mathbf{r}))^2 d\mathbf{r}}$, was used.

Once the distances between the relaxed images were computed, an interpolation scheme was used to determine new, evenly spaced images that were then the starting point for the next iteration of the algorithm, which continued until convergence was achieved. Further details are given in the Supplementary Text.

## Open versus closed systems

Physically, nucleation of a dense phase from a weak phase is very different in finite volumes for the cases of fixed and variable numbers of particles because, in the former case, the ambient density can drop substantially as the material is incorporated into the dense phase. The present approach can be applied to either case. The deterministic dynamics (Eq. 5) conserves particle number except at the boundary, where the behavior depends on the applied boundary conditions. In general, the interpolation step may or may not preserve particle number depending on the implementation used, but here, to be applicable to both open and closed systems, a conservative interpolation scheme was used. In this study, attention has been restricted to the case of an open system (at fixed chemical potential) as this is the more generic case (the results for finite systems vary according to their volume). In principle, the open system—even with a finite computational cell—should give results close to what one would obtain for an infinite system (for which the fixed/variable particle number distinction becomes irrelevant). Technical details can be found in the Supplementary Text.

## SUPPLEMENTARY MATERIALS

Supplementary material for this article is available at http://advances.sciencemag.org/cgi/content/full/5/4/eaav7399/DC1

Supplementary Text
Fig. S1. Phase diagram as presented in (50).
Fig. S2. Supersaturation as a function of the excess number of particles for liquid critical droplets.
Fig. S3. Initial and final paths for crystallization with a final cluster of 286 molecules using 50 images and a final path using 20 images showing the excess free energy as a function of Euclidean distance along the path (arbitrarily scaled so that the distance between images is equal to one).
Fig. S4. Log of the density for slices of images from the initial guess to the pathway.
Fig. S5. Nucleation pathways for several systems displayed in terms of the free energy along the paths.
Fig. S6. Nucleation pathways for several systems displayed in terms of the number of molecules in the clusters along the paths.
Fig. S7. Excess number of particles in the critical clusters as a function of supersaturation.
Fig. S8. Excess free energies in the critical clusters as a function of supersaturation.
Table S1. Thermodynamic quantities at weak-solution/dense-solution liquid-liquid coexistence.
Table S2. Thermodynamic quantities used in liquid-liquid calculations.
Movie S1. Cross section of liquid droplet density as it evolves along the nucleation pathway for the example presented in the main text showing the log of the density.
Movie S2. Cross section of liquid droplet density as it evolves along the nucleation pathway for the example presented in the main text showing the density on a linear scale.
Movie S3. Cross section of solid cluster density as it evolves along the nucleation pathway for the example presented in the main text showing the log of the density.
Movie S4. Cross section of solid cluster density as it evolves along the nucleation pathway for the example presented in the main text showing the density on a linear scale.
References (62–65)

**Acknowledgments:** I thank D. Maes and J. Lam for helpful comments concerning the manuscript. **Funding:** This work was supported by the European Space Agency (ESA) and the Belgian Federal Science Policy Office (BELSPO) in the framework of the PRODEX Programme (contract number ESA AO-2004-070). Computational resources have been provided by the Shared ICT Services Centre, Université Libre de Bruxelles. **Author contributions:** All work was performed by the author. **Competing interests:** The author declares that he has no competing interests. **Data and materials availability:** All codes used in this work are available from the author's open source project found at https://jimlutsko.github.io/classicalDFT. Additional data related to this paper may be requested from the author.

Submitted 15 October 2018
Accepted 12 February 2019
Published 5 April 2019
10.1126/sciadv.aav7399

Citation: J. F. Lutsko, How crystals form: A theory of nucleation pathways. *Sci. Adv.* **5**, eaav7399 (2019).

# Science Advances

**How crystals form: A theory of nucleation pathways**
James F. Lutsko